\documentstyle[eqsecnum,aps]{revtex}

\input epsf.sty

\newcommand{\lci}{i}

\makeatletter
\@addtoreset{equation}{section}
\makeatother

\newcommand{\ps}{\psi}

\newcommand{\psb}{\overline{\ps}}

\newcommand{\be}{\begin{equation}}
\newcommand{\ee}{\end{equation}}
\newcommand{\bea}{\begin{eqnarray}}
\newcommand{\eea}{\end{eqnarray}}
\newcommand{\eq}{\ref}
\newcommand{\beq}{\begin{equation}}
\newcommand{\eeq}{\end{equation}}

\newcommand{\lb}{\label}

\newcommand{\lsim}{\stackrel{<}{\sim}} 

\def\hs{{\hat s}}
\def \3{\ss}

\begin{document}
%
%
\twocolumn[\hsize\textwidth\columnwidth\hsize\csname%
@twocolumnfalse\endcsname
\flushright{UAB-FT-470} 

\title{The $7/11$ rule: an estimate of $m_\rho/f_\pi$}

\author{Maarten F.L. Golterman}
\address{Grup de Fisica Teorica and IFAE,
Universitat Autonoma de Barcelona, \\
08193 Bellaterra (Barcelona), Spain \\
and \\
Department of Physics, Washington University, \\
St. Louis, MO 63130, USA\cite{pa}
}
\author{Santiago Peris}
\address{Grup de Fisica Teorica and IFAE, 
Universitat Autonoma de Barcelona, \\
08193 Bellaterra (Barcelona), Spain
}
\maketitle
\begin{abstract}
We derive an estimate for the ratio of the rho mass and the 
pion decay constant from an analysis of vector and axial-vector
two-point functions using large-$N_c$, lowest-meson dominance and
the operator product expansion, in the chiral limit.  
We discuss the extension of this 
analysis to the scalar and pseudoscalar sector. Furthermore, this leads to a
successful parameter-free determination of the $L_i$ couplings of the
chiral Lagrangian if an improved Nambu-Jona-Lasinio {\it ansatz} for Green
functions is assumed at low energies.
\end{abstract}
]                                                    
\narrowtext
\section{Introduction}
\label{SEC1}
The large-$N_c$ expansion \cite{`t Hooft} offers at present the only
nonperturbative analytic approximation to QCD. This approximation is powerful 
enough to allow, for instance, a proof (within certain assumptions) of the
expected pattern of chiral symmetry breaking \cite{Coleman-Witten}. In
particular, for Green functions of color-singlet quark bilinears,
the lowest order in this expansion corresponds to a theory of an infinite
number of stable mesons. 

At low energies there exists an effective field theory description 
in terms of what is called Chiral Perturbation
Theory\cite{Weinberg,GL} which corresponds to a systematic expansion 
of Green functions involving Goldstone mesons in
powers of momenta and light quark masses. The resonance saturation of the 
$O(p^4)$ Gasser--Leutwyler low-energy constants $L_i$ of Refs.~\cite{SC1,SC2} 
can then be viewed as the matching between Chiral Perturbation Theory and
large-$N_c$ QCD in the low-energy regime~\cite{Peris-Perrottet-deRafael}.

At high energies, QCD is described more efficiently by weak-coupling
perturbation theory (enhanced with the Operator Product Expansion
(OPE)). Unlike the 
low-energy counterpart, the matching between the high-energy regimes of 
large-$N_c$ QCD and perturbative QCD has been studied much less
systematically, even though it has an enormous impact on our understanding of
important problems such as weak matrix elements, see for
instance Ref.~\cite{Knecht-Peris-deRafael}.

Following Ref.~\cite{Peris-Perrottet-deRafael} we will 
assume that the spectrum of this infinite tower of mesons  
can be reasonably approximated by just one resonance plus  
a sharp onset of the perturbative
continuum at a scale $s_0$ 
(global duality) \cite{SVZ,EDRreview}.  Furthermore, in order to
simplify the analysis, we will work in the chiral limit.
The scale $s_0$ is fixed by the requirement of matching to the first 
term in the OPE\cite{BLdeR}. Higher terms in the OPE 
then become fixed \cite{Knecht-deRafael} in terms of the resonance
parameters. 

Concerning the reliability of this description we 
expect a certain ``hierarchy" in this approach. First, at higher orders 
in the OPE, our
simple description is likely to break down, but then the relative importance 
of the corresponding terms is also suppressed by higher powers of momenta. 
In practice, how far one can go is an
open question, and may depend on the channel one is looking at. 
Second, one should expect that any linear combination of sum rules
that is an order parameter of chiral symmetry ({\it i.e.} 
which would vanish were chiral symmetry unbroken) would be better
approximated by our one-resonance-plus-continuum description than something
that is not, as the former does not depend on our simple {\it ansatz} 
for the onset of the perturbative
continuum, since it must cancel by definition in such combinations. 

In Sect. II, we will consider two-point functions of the vector and
axial-vector (nonsinglet) flavor currents.  This will lead us to a new
relation between the rho mass $m_\rho$, its electromagnetic decay constant
$f_\rho$, and $f_\pi$.  This relation turns out to work 
remarkably well when compared to experimental values of these parameters.
In Sect. III, we discuss the extension of this type of analysis to 
the scalar and pseudo-scalar sector.   While we find a relatively stable
value for the scalar mass, this extension also exhibits the limitations
of the approach, notably with respect to the chiral condensate.
We make contact with a previous analysis based on a Nambu--Jona-Lasinio
{\it ansatz} in Sect. IV, and summarize in Sect. V. 

\section{Vectors}
\label{SEC2}

Although historically the two Weinberg sum rules \cite{Wein}
for the $\rho$ and $a_1$ mesons, 
\bea
&&f_\rho^2m_\rho^2=f_{a_1}^2m_{a_1}^2+f_\pi^2\;, \lb{WEIN1} \\
&&f_\rho^2m_\rho^4=f_{a_1}^2m_{a_1}^4\;, \lb{WEIN2}
\eea
came before the large-$N_c$ expansion, they can be derived in 
our context in the following way~\cite{EDRreview}.  
One assumes that the vector-current
two-point function $\Pi_V(q^2)$, defined from
\bea
\Pi^V_{\mu\nu}(q)&=&i\int dx\;e^{iqx}\langle T(V_\mu(x)V^\dagger_\nu(0))\rangle 
\lb{PIV} \\
&=&(q_\mu q_\nu-g_{\mu\nu}q^2)\Pi_V(q^2)\;, \nonumber
\eea
with $V_\mu(x)={\bar d}(x)\gamma_\mu u(x)$, can be described by (up to one
subtraction)
\bea
\Pi_V(Q^2)&=&\int_0^\infty\frac{dt}{t+Q^2}\frac{1}{\pi}{\rm Im}\;\Pi_V(t)\;,
\nonumber \\
\frac{1}{\pi}{\rm Im}\;\Pi_V(t)&=&2f_\rho^2m_\rho^2\delta(t-m_\rho^2) \lb{IMPIV} \\
&&+\left(\frac{4}{3}\frac{N_c}{(4\pi)^2}+\dots\right)\theta(t-s_0)\;, \nonumber
\eea 
where $f_\rho$ is the electromagnetic decay constant of the rho.
The ellipsis represents higher-order perturbative corrections 
above a certain energy $\sqrt{s_0}$. 

In the chiral limit, an analogous {\it ansatz} then holds for ${\rm Im}\;\Pi_A$
in terms of corresponding parameters for the $a_1$,
defined from the axial current $A_\mu(x)={\bar d}(x)\gamma_\mu\gamma_5 u(x)$,
except that a term representing the pion, $2f_\pi^2(t)\delta(t)$, 
has to be added (we use the convention in which the experimental value of 
$f_\pi\simeq 93$~MeV).  
Since perturbative QCD does not see chiral symmetry breaking, one takes $s_0$
the same in the vector and axial channels.      

One may now calculate $\Pi_{V,A}(Q^2)$ for Euclidean 
momentum $Q$ from Eq.~(\eq{IMPIV})
and confront it with the OPE-result for large $Q^2$.  One obtains the following
set of relations \cite{SVZ,BLdeR,Peris-Perrottet-deRafael}:
\bea
2f_\rho^2m_\rho^2-\frac{4}{3}\frac{N_c}{(4\pi)^2}s_0&=&0\;, \lb{RHO2} \\
-2f_\rho^2m_\rho^4+\frac{2}{3}\frac{N_c}{(4\pi)^2}s_0^2&=&
\frac{1}{12\pi}\langle\alpha_s G_{\mu\nu}G^{\mu\nu}\rangle\;, \lb{RHO4} \\
2f_\rho^2m_\rho^6-\frac{4}{9}\frac{N_c}{(4\pi)^2}s_0^3&=&
-\frac{28}{9}\pi\alpha_s\langle\psb\psi\rangle^2\;, \lb{RHO6} 
\eea
where the second term on the left-hand side of each of these equations comes
from the leading term in the perturbative contribution to ${\rm Im}\;\Pi_V$,
and the right-hand side is leading order in $\alpha_s$ and $1/N_c$.
(Note that all terms are of the same order in $N_c$.)
For the axial channel, one obtains
\bea
2f_{a_1}^2m_{a_1}^2+2f_\pi^2-\frac{4}{3}\frac{N_c}{(4\pi)^2}s_0&=&0\;, \lb{A2} \\
-2f_{a_1}^2m_{a_1}^4+\frac{2}{3}\frac{N_c}{(4\pi)^2}s_0^2&=&
\frac{1}{12\pi}\langle\alpha_s G_{\mu\nu}G^{\mu\nu}\rangle\;, \lb{A4} \\
2f_{a_1}^2m_{a_1}^6-\frac{4}{9}\frac{N_c}{(4\pi)^2}s_0^3&=&
\frac{44}{9}\pi\alpha_s\langle\psb\psi\rangle^2\;. \lb{A6}
\eea
Combining Eqs.~(\eq{RHO2},\eq{A2}) and eliminating the gluon condensate between
Eqs.~(\eq{RHO4},\eq{A4}) yields the
two Weinberg sum rules in the form Eqs.~(\eq{WEIN1},\eq{WEIN2}).

One may expect these sum rules to be reasonably well
satisfied, if the ``onset" of perturbation theory 
occurs at a scale $\sqrt{s_0}$
(for which a phenomenological value can be calculated from Eq.~(\eq{RHO2}))
above $m_{a_1}$, where presumably one may trust 
perturbative QCD.  In addition, the sum rules do not
depend on the gluon condensate.
Eq.~(\eq{RHO4}) then gives a phenomenological
estimate for the gluon condensate as well, but
this estimate may be less reliable, because it arises as the difference between
two large numbers, one representing the low-energy, and the other representing 
the high-energy behavior of $\Pi_V$.

So far, we only reviewed a derivation of long-known results.  However,
we may carry this procedure one step further, and also eliminate the fermion
condensate between Eqs.~(\eq{RHO6},\eq{A6}). Substituting the value of $s_0$
from Eq.~(\eq{RHO2}), the result is a relation between $m_\rho$, $f_\rho$
and $f_\pi$:
\be
\frac{1-\frac{3}{4}\frac{(4\pi)^4}{N_c^2}f_\rho^4}
{(1-\frac{f_\pi^2}{f_\rho^2m_\rho^2})^{-1}-
\frac{3}{4}\frac{(4\pi)^4}{N_c^2}f_\rho^4}
=-\frac{7}{11}\;,\lb{SEVENELEVEN}
\ee
once $m_{a_1}$ and $f_{a_1}$ are eliminated using Weinberg's sum rules.

The fact that the right-hand side of Eq.~(\eq{SEVENELEVEN}) is negative 
already leads to an interesting lower bound on $f_\rho$.  First, note that
$f_\rho^2m_\rho^2>f_\pi^2$, from Eq.~(\eq{WEIN1}).  It then follows that
the denominator of the left-hand side of Eq.~(\eq{SEVENELEVEN}) is larger
than the numerator, and hence that the numerator has to be negative.  In
fact, we have
\be
\frac{4}{3}\frac{N_c^2}{(4\pi)^4}<f_\rho^4<\frac{4}{3}\frac{N_c^2}{(4\pi)^4}
\frac{1}{1-\frac{f_\pi^2}{f_\rho^2m_\rho^2}}\;,\lb{INEQ}
\ee
and from the left inequality it follows that $f_\rho>0.148$.  Using
$f_\pi=87$~MeV (as an estimate of what its value would be in the
chiral limit \cite{GL}), and $m_\rho=770$~MeV, the right inequality
leads to $f_\rho<0.171$.  This compares well with the experimental
value $f_\rho=0.20$ if we take into account that
we have used the large-$N_c$ (narrow resonance) and chiral approximations.

At this point we bring in yet another ingredient.  Within our set of 
assumptions, another relation between  $m_\rho$, $f_\rho$
and $f_\pi$ was previously obtained in Ref. \cite{SC1} 
from the requirement that the pion electromagnetic form factor and
axial form factor in $\pi \rightarrow e \nu_e \gamma$ satisfy
unsubtracted dispersion relations: 
\be
f_\rho^2m_\rho^2=2f_\pi^2\;. \lb{FF}
\ee
The right inequality now translates into $f_\rho<0.176$ (12\% below the
experimental value).  In addition,
this leads to a complete solution for all the parameters in the vector and
axial channels in terms of $f_\pi$ (for $N_c=3$):
\bea
&&(4\pi f_\rho)^2=4(4\pi f_{a_1})^2= \frac{10}{\sqrt{6}}\;, \lb{FS} \\
&&m_\rho^2=m_{a_1}^2/2= \frac{\sqrt{6}}{5}(4\pi f_\pi)^2\;, \lb{MS} 
\eea
and
\bea
\sqrt{s_0}&=&4\pi f_\pi\;, \lb{S0} \\
\langle\alpha_s G_{\mu\nu}G^{\mu\nu}\rangle&=&\frac{384}{5}\pi^3
(5-2\sqrt{6})f_\pi^4\;, \lb{GG} \\
\pi\alpha_s\langle\psb\psi\rangle^2&=&\frac{768}{25}\pi^4f_\pi^6\;. \lb{PP}
\eea
Using $f_\pi=87$~MeV, this gives $m_\rho=765$~MeV, $m_{a_1}=1082$~MeV, 
$f_\rho=0.16$ and $f_{a_1}=0.08$ which are remarkably good ($f_{a_1}$
is only poorly known experimentally). 

The scale $\sqrt{s_0}=1.09$~GeV, slightly larger than the
$a_1$ mass, is acceptable for the ``onset" of perturbation theory.  Even
the condensates (which, as argued above, might be expected to fare less
well) are reasonable: $\langle\alpha_s G_{\mu\nu}G^{\mu\nu}\rangle=
0.014$~GeV$^4$, and $\pi\alpha_s\langle\psb\psi\rangle^2=
(330~{\rm MeV})^6\simeq 13 \times 10^{-4}$~GeV$^6$. 
These numbers may for instance be compared to 
$\langle\alpha_s G_{\mu\nu}G^{\mu\nu}\rangle\simeq 0.048 \pm 0.030$~GeV$^4$
from Ref. \cite{Yndurain}, and to the {\it combination} 
$\pi\alpha_s\langle\psb\psi\rangle^2\simeq (9 \pm 2)\times 10^{-4}$~GeV$^6$
as obtained from the fit to tau decays with the four-quark condensate 
recently performed in Ref. \cite{Girlanda}.

Using the two-loop running of $\alpha_s$ in the $\overline{MS}$ scheme,
\be
\frac{\alpha_s(\mu)}{\pi}=\frac{A}{\log{\mu^2/\Lambda^2}}
\left(1-B\frac{\log{\log{\mu^2/\Lambda^2}}}{\log{\mu^2/\Lambda^2}}
\right)\;,
\lb{RUN}
\ee
with $A=4/11$ and $B=102/121$ for $N_c=\infty$ and $A=4/9$, $B=64/81$
at $N_c=n_f=3$, one can now try to give an estimate for the value of the 
chiral condensate
at the scale $s_0$.  Using $\Lambda=300$--$400$~MeV and $N_c=3,\infty$,
with $\alpha_s(\sqrt{s_0})/\pi=0.097$--$0.160$, one obtains
\be
\langle\psb\psi\rangle(\sqrt{s_0})=-(319\pm 13~{\rm MeV})^3\;, \lb{COND}
\ee
where the error comes from the variation of $\alpha_s$.

At this point, we would like to make several comments about the
stability of these results.

First, unlike the Weinberg sum rules, Eq.~(\eq{SEVENELEVEN})
does depend on higher-order corrections in $\alpha_s$ to the $s_0^n$
terms and the condensate terms in Eqs.~(\eq{RHO2}--\eq{A6}).  To leading
order, $\alpha_s$-corrections to the $s_0^n$ terms can be incorporated
by replacing $N_c/(4\pi)^2\to N_c(1+\alpha_s/\pi)/(4\pi)^2$ in 
Eq.~(\eq{SEVENELEVEN}) \cite{Kataev}, 
which leads to a value of $f_\rho$ increased
by a factor $1+\alpha_s/(2\pi)$.  With the range of values for $\alpha_s$
above, this leads to an increase of at most 8\%, or a decrease of $m_\rho$
by the same amount.  There are also perturbative corrections to the
ratio $-7/11$ in Eq.~(\eq{SEVENELEVEN}), which are not known to us. 
(There is no contribution from dimension 6 gluon condensates to the order 
considered\cite{HM}.) However,  
recent attempts to extract information on the actual value of this
ratio from $\tau$ decays data are consistent
with a value not very different from $-7/11$ \cite{PICH}.  A change of
the ratio $-7/11$ by 10\% leads to a change of $f_\rho$ of about 1\%.
Note also that just the fact that the right-hand side of Eq.~(\eq{SEVENELEVEN})
is negative already led to an estimate $f_\rho=0.16(1)$, given the
experimental value of the rho mass.

 Second, one expects the estimates for the condensates to be less reliable
than those for the $\rho$ and $a_1$ parameters as they 
arise from a subtle balance
between the low- and high-energy {\it ans{\"a}tze} for the vector and
axial-vector two-point functions.  We also
remind the reader that, as a matter of
principle, the gluon condensate has a physical meaning only in
conjunction with the perturbative part and not in isolation~\cite{MS}.
All this, however, does not affect the
Weinberg sum rules or the sum rule Eq.~(\eq{SEVENELEVEN}).

Third, one may expect that chiral corrections to the ratio $m_\rho/f_\pi$
are not large by noticing that the experimental value, 8.3, is close
to $m_{K^*}/f_K=7.9$.  (The difference between the two ratios should
be due mainly to chiral corrections from the strange quark mass.)

\section{Scalars}
\label{SEC3}
In the previous section, we have applied some rather simple
phenomenological assumptions to an analysis of vector and
axial-vector two-point functions.  This led to a complete and
rather remarkable determination of the vector and axial-vector
resonance parameters $m_{\rho,a_1}$ and $f_{\rho,a_1}$.  Here we
would like to explore what happens when we attempt to
do the same in the (pseudo)scalar sector.

Again, we assume that, in the large $N_c$ and chiral limits, the
scalar and pseudoscalar two-point functions, given by 
\be
\Pi_{S,P}(q)=i\int dx\;e^{iqx}\langle T(J_{S,P}(x)J_{S,P}^\dagger(0))
\rangle\;,
\lb{SPS}
\ee
where $J_S((x)={\overline d}(x)u(x)$ and $J_P(x)={\overline d}(x)
\gamma_5 u(x)$, can be described by
\be
\Pi_{S,P}(Q^2)=\int_0^\infty\frac{dt}{t+Q^2}\frac{1}{\pi}{\rm Im}\;
\Pi_{S,P}(t)+{\rm subtractions}\;,\lb{DISP}
\ee
with
\bea
{\rm Im}\;\Pi_S(t)&=&16B^2c_m^2\delta(t-m_S^2)+\frac{N_c}{(4\pi)^2}
\kappa 2t\theta(t-\hs_0)\;,\nonumber\\
{\rm Im}\;\Pi_P(t)&=&2B^2f_\pi^2\delta(t)+
16B^2d_m^2\delta(t-m_P^2) \lb{IMPISP} \\
&&+\frac{N_c}{(4\pi)^2}
\kappa 2t\theta(t-\hs_0)\;,\nonumber
\eea
where we use the notation of Ref.~\cite{SC2}.  The scale $\hs_0$
is not necessarily the same as the scale $s_0$ in the vector sector.
To first order in
perturbation theory, the factor $\kappa$ is known to contain very large
$\alpha_s$ corrections\cite{BPR}
\be
\kappa=1+\frac{\alpha_s(\mu)}{\pi}\left(\frac{17}{3}
-2\log{\frac{t}{\mu^2}}\right)\;,\lb{KAPPA}
\ee
which is the reason why we do not discard it right from the beginning. 
Eq.~(\eq{KAPPA}) is only valid at $N_c=3$. However, as we will see in the
numerical analysis that follows, these large $\alpha_s$ corrections 
turn out to have no significant impact on the numerical results. 

For large Euclidean $Q^2$, we may again confront $\Pi_{S,P}(Q^2)$ 
calculated
from Eqs.~(\eq{DISP}-\eq{KAPPA}) with the OPE result of
Ref.~\cite{JM}. Taking $\mu=\sqrt{\hs_0}$, and equating inverse powers
of $Q^2$, one finds, with $B= - \langle\psb\psi\rangle/f_\pi^2$, in the
large $N_c$ and chiral limits
\bea
16B^2c_m^2-\frac{N_c}{16\pi^2}\hs_0^2\kappa_1(\hs_0)&=&
\frac{1}{8\pi}\langle\alpha_s G_{\mu\nu}G^{\mu\nu}\rangle\;,\lb{S1}\\
-16B^2c_m^2m_S^2+\frac{N_c}{24\pi^2}\hs_0^3\kappa_2(\hs_0)&=&
-\frac{22}{3}\pi\alpha_s\langle\psb\psi\rangle^2 \lb{S2}
\eea
in the scalar sector, and
\bea
2B^2f_\pi^2+16B^2d_m^2-\frac{N_c}{16\pi^2}\hs_0^2\kappa_1(\hs_0)=
\lb{P1}\\
&\!\!\!\!\!\!\!\!\!\!\!\!\!
\frac{1}{8\pi}\langle\alpha_s G_{\mu\nu}G^{\mu\nu}\rangle\;,\nonumber\\
-16B^2d_m^2m_P^2+\frac{N_c}{24\pi^2}\hs_0^3\kappa_2(\hs_0)=\lb{P2}\\
&\!\!\!\!\!\!\!\!\!\!\!\!\!
\frac{14}{3}\pi\alpha_s\langle\psb\psi\rangle^2 \nonumber
\eea
in the pseudoscalar sector.  The leading perturbative corrections $\kappa_1$
and $\kappa_2$ stem from the $\kappa$ factor in Eq. (\eq{KAPPA}), and read
\bea
\kappa_1(\hs_0)&=&1+\frac{20}{3}\frac{\alpha_s(\sqrt{\hs_0})}{\pi}\;,
\lb{KAPPAS}\\
\kappa_2(\hs_0)&=&1+\frac{19}{3}\frac{\alpha_s(\sqrt{\hs_0})}{\pi}\;. 
\eea

At this point, let us discuss how we may try to explore these equations.
In order to eliminate the dependence on the gluon condensate
and the perturbative terms, we might choose to first consider combinations
that are order parameters of chiral symmetry, and look at the
differences (\eq{S1})$-$(\eq{P1}) and (\eq{S2})$-$(\eq{P2}), {\it
  i.e.}~\cite{Marc-Eduardo,Andrianov}: 
\bea
8c_m^2-8d_m^2-f_\pi^2&=&0\;,\lb{S1P1}\\
B^2c_m^2m_S^2-B^2d_m^2m_P^2&=&{3\over 4} \pi\alpha_s\langle\psb\psi\rangle^2\;.
\lb{S2P2}
\eea
Given the values of $\pi\alpha_s\langle\psb\psi\rangle^2$ (for instance
from the vector sector, {\it cf.} Eq.~(\eq{PP})) and $m_{S,P}$ we may
solve these equations for $c_m$, $d_m$, which are then still logarithmically
dependent on the scale $\hs_0$ (through $\alpha_s$).  
The additional equation is provided
by Eq.~(\eq{S1}) (or equivalently Eq.~(\eq{P1})), 
and we obtain a solution for $c_m$, $d_m$ and $\hs_0$.
It turns out that the solution is very insensitive to the gluon 
condensate (see below).
This leaves us with the question of what values to use for $m_{S,P}$.
We will assume that $m_P$ corresponds to the mass of the $\pi(1300)$, which is
firmly established in the Particle Data Tables.
As is well known, the situation in the scalar sector 
is much less understood (for a recent discussion, see Ref.~\cite{PENN}).
Therefore, instead of directly guessing a 
value for the scalar mass, we will add an extra
equation.  Within our set of assumptions, the low-energy constant
$L_8$ \cite{GL} is determined by \cite{SC2}
\be
L_8=\frac{c_m^2}{2m_S^2}-\frac{d_m^2}{2m_P^2}\;.\lb{L8}
\ee
The experimental value is $L_8=0.9(3)\times 10^{-3}$ \cite{Li}.

Note that we have not used Eq.~(\eq{S2}) (or, equivalently,
Eq.~(\eq{P2})) separately yet.  

Let us now consider solutions.  Taking $f_\pi=87$~MeV, 
$\Lambda_{\overline{MS}}=372$~MeV \cite{LAMBDA}, 
$m_P=1300$~MeV, and $L_8=0.0009$,
and using the results Eqs.~(\eq{GG},\eq{PP}) for the condensates,
we obtain
\bea
&c_m=41~{\rm MeV}, &d_m=27~{\rm MeV},\lb{CD1}\\
&m_S=0.86~{\rm GeV}, &\sqrt{\hs_0}=2.16~{\rm GeV}\;.\lb{MASS1}
\eea
Note that the $\pi$(1300) ``decouples from" $L_8$ in that it only 
gives a very small contribution to it. 

Except for $\hs_0$, these results are rather stable. In
order to demonstrate this, we will vary one input at a time and see how
the values of $c_m$, $d_m$, $m_S$ and $\hs_0$ change.

Varying the gluon condensate from 0 to 10 times the value
given in Eq.~(\eq{GG}) has no significant impact on the solution.  
Changing $\Lambda_{\overline{MS}}$
to 300~MeV changes these numbers by at most about 5\%. Even changing the
$\alpha_s$ corrections in the $\kappa$ factor by a huge factor such as 10,
or omitting the $\alpha_s$ correction altogether, does
not significantly change these numbers (except the value of $\hs_0$).

Changing $m_P$, and even more so $L_8$, has a bigger 
effect (note that these effects
are related through Eq.~(\eq{L8})).  Taking $m_P$ from
1400~MeV to 1200~MeV (keeping $L_8=0.0009$) leads to
\bea
38~{\rm MeV}<&c_m&<46~{\rm MeV}\;,\lb{RANGEMP}\\
22~{\rm MeV}<&d_m&<34~{\rm MeV}\;,\nonumber\\
0.84~{\rm GeV}<&m_S&<0.90~{\rm GeV}\;,\nonumber\\
2.08~{\rm GeV}<&\sqrt{\hs_0}&<2.32~{\rm GeV}\;.\nonumber
\eea
We found no solution for $m_P\lsim 1$GeV.
Similarly, changing $L_8$ from 0.0012 to 0.0006 (keeping $m_P=1300$~MeV)
gives
\bea
36~{\rm MeV}<&c_m&<81~{\rm MeV}\;,\lb{RANGEL8}\\
19~{\rm MeV}<&d_m&<74~{\rm MeV}\;,\nonumber\\
0.71~{\rm GeV}<&m_S&<1.20~{\rm GeV}\;,\nonumber\\
2.02~{\rm GeV}<&\sqrt{\hs_0}&<3.25~{\rm GeV}\;.\nonumber
\eea

Although all these solutions satisfy the difference 
(\eq{S2})$-$(\eq{P2}), they do
not satisfy Eq. (\eq{S2}) or Eq. (\eq{P2}) separately.
With the value of $\pi\alpha_s\langle\psb\psi\rangle^2$ from the
vector sector, there does not appear to exist any reasonable solution
to the four equations
Eqs.~(\eq{S1}-\eq{P2}) (for any reasonable choice of $m_P$).
Perhaps another indication of this problem is the value of $\hs_0$, which
is substantially larger than the value of $s_0$ found in the vector
sector.  At $\Lambda_{\overline{MS}}=372$~MeV, one has
$\alpha_s(\hs_0)\approx 0.09(0.07)$, to be compared with
$\alpha_s(s_0)\approx 0.15(0.12)$ for $N_c=n_f=3$ ($N_c=\infty$).
This makes a difference of about 20\% in the value of $\langle
\psb\psi\rangle$.  If we solve the same set of equations, but with
Eq.~(\eq{S2}) replacing Eq.~(\eq{S1}), we obtain, instead of
the result (\eq{CD1}) and (\eq{MASS1}),
\bea
&c_m=40~{\rm MeV}, &d_m=26~{\rm MeV},\lb{CD2}\\
&m_S=0.86~{\rm GeV}, &\sqrt{\hs_0}=1.64~{\rm GeV}\;.\lb{MASS2}
\eea
Resonance parameters stay the same; the only thing that changes is 
the value of $\hs_0$. This is an example of a point  we made in the
Introduction: our simple {\it ansatz} for the onset of the perturbative
continuum seems to make
Eq.~(\eq{S2}) and Eq.~(\eq{P2}) incompatible (but, see below).  

However, since the continuum
cancels in the difference, the combination (\eq{S2})$-$(\eq{P2}) 
does lead to a solution. Of course, the lack of a solution to both 
Eqs.~(\eq{S2}) and Eq.~(\eq{P2}) may merely mean that our spectrum is too
simple to reproduce the physics of dimension-six operators in the OPE.
However, even in the present case, we may appreciate how stable resonance
parameters seem to be by means of the following little exercise. First notice
that, up to now, we have been using the value of $\pi\alpha_s\langle
\psb\psi\rangle^2$ obtained in the (axial)vector sector. Just as we studied
the effect of the variation of other inputs earlier, 
this brings us now to the sensitivity of the solution in the 
scalar-pseudoscalar sector to the value of the quark condensate. 
It is interesting to see  what happens if we lower, {\it by fiat}, 
the value of this four-fermion condensate.  (The value of 
$\langle \psb\psi\rangle$ in Eq.~(\eq{COND}) is rather
high in comparison with other phenomenological estimates.)  As it
turns out, we happen to find a
solution for $m_{S,P}$, $c_m$, $d_m$ and $\hs_0$ from the whole set 
of equations (Eqs.~(\eq{S1}-\eq{P2},\eq{L8})) if we lower
the value of Eq.~(\eq{PP}) by a fudge factor 11:
\bea
&c_m=51~{\rm MeV}, &d_m=41~{\rm MeV},\lb{CD3}\\
&m_S=0.92~{\rm GeV}, &m_P=1.13~{\rm GeV},\lb{MASS3}\\
&\sqrt{\hs_0}=1.14~{\rm GeV}\;.&\null \nonumber
\eea
(For reduction factors smaller than 11, we find $\hs_0<m_P^2$, which is not
acceptable: the basic assumption was that two-point functions
can be described by a small number of narrow resonances below
an energy scale $s_0$, and perturbative QCD above that scale.)
What we learn is that even in this case the solution 
(in particular the scalar mass) stays roughly
in the neighborhood of the resonance parameters found 
in Eqs.~(\eq{CD1},\eq{MASS1}).

While it is clear that the scalar sector is less simple than
the vector sector in this approach, one interesting fact
emerges from this section: it seems likely that there exists 
a scalar resonance with mass between 700 and 1200~MeV
in the large-$N_c$ and chiral limits.

\section{The $L_\lci$ couplings}

In Ref. \cite{Peris-Perrottet-deRafael} it was shown how an extended 
Nambu--Jona-Lasinio~\cite{Hans} 
{\it ansatz} for Green's functions in QCD
can be improved to restrict the parameters  of the 
lowest-meson dominance approximation to the 
large-$N_c$ limit in a way which is compatible with several examples 
of the OPE. The final outcome of this analysis was a 
determination of all the $L_i$'s leading at large $N_c$ 
in terms of the ratio $f_{\pi}/m_{\rho}$. At $N_c=3$, they read  
\cite{Peris-Perrottet-deRafael}:

\bea
6 L_1 =& 3 L_2 = - \frac{8}{7} L_3 = 4 L_5 = 8 L_8 = \frac{3}{4} L_9 = -
L_{10}\nonumber\\ 
&=\frac{3}{8} \frac{f_{\pi}^2}{m_{\rho}^2}\;.\lb{Li}
\eea
Since we can now use Eq. (\eq{MS}) to fix this ratio to be 
\be
\frac{3}{8}\frac{f_{\pi}^2}{m_{\rho}^2} = \frac{15}{8 \sqrt{6}} \frac{1}{16
  \pi^2}\; , \lb{end}
\ee
Eqs.~(\eq{Li},\eq{end}) lead to a parameter-free determination for the
$L_i$ couplings.

\begin{figure}
\def\filename{Li.ps}
\begin{center}
\epsfxsize=9.30cm
\epsfbox{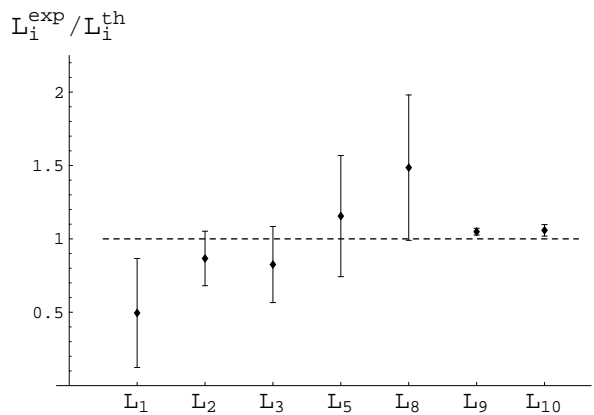}
\end{center}
\caption{Ratios of the experimental values and 
those obtained from Eq. ~(\eq{Li},\eq{end}) for the $O(p^4)$ constants $L_i$.  
The error bars represent the experimental errors.}
\label{fig:Lis}
\vspace*{1cm}
\end{figure}

This is plotted in Fig.~\ref{fig:Lis}, where we use $\mu=m_{\rho}$ for the
renormalization scale in these couplings to estimate their experimental
values.  The error bars represent only the errors in the experimental values,
and one should keep the uncertainty in the choice of renormalization
scale in mind while considering Fig.~\ref{fig:Lis}.  For instance, if we change
the scale from $m_\rho$ to 1~GeV, the couplings $L_9$ and $L_{10}$ 
change by $-6$\%, resp. 8\%,
which is more than the experimental error.  Of course,
a more precise knowledge about the scale $\mu$ at which the matching
takes place would be required for a more detailed analysis (which
would necessarily go beyond leading order in $1/N_c$).

\section{Discussion}
\label{SEC4}

Let us summarize what we have learned from this simple large-$N_c$
inspired model for vector and scalar two-point functions.

First, we showed that the usual ``large-$N_c$ plus
lowest-meson dominance" derivation of the
two Weinberg sum rules can be extended to give us a third sum rule,
Eq.~(\eq{SEVENELEVEN}), relating vector masses and couplings.
Like the first two sum rules, this sum rule is nontrivial because
of the fact that chiral symmetry is spontaneously broken,
with nonvanishing $f_\pi$ and $\langle\psb\psi\rangle$.

This new sum rule can be combined with Eq.~(\eq{FF}) to express
$m_\rho$, $m_{a_1}$ and $f_\rho$, $f_{a_1}$ in terms of $f_\pi$.
We find $m_\rho/f_\pi=8.8$, and $f_\rho=0.16$, to be compared
with the experimental values $m_\rho/f_\pi=8.3$, and $f_\rho=0.20$.  
(The $a_1$
parameters then follow from the Weinberg sum rules in the usual way.)
The agreement between our values and the experimental ones is
very good, considering that the $1/N_c$ and chiral expansions
are ingredients of our analysis.
 
The scale for the onset of perturbation theory comes
out slightly higher than the $a_1$ mass.  One also obtains
values for the gluon and chiral condensates.  Note, however,
that the three sum rules are derived by eliminating the condensates,
and that our results for $m_\rho$ and $f_\rho$ are therefore 
independent of their values. 

Then, we presented a similar analysis for the scalar and
pseudoscalar two-point functions.  An important qualitative
difference is the fact that, in this case, restricting 
ourselves to the same set of condensates in the OPE, only
four (Eqs.~(\eq{S1}-\eq{P2})) instead of six 
(Eqs.~(\eq{RHO2}-\eq{A6}) are obtained.  This is related to
the fact that in the scalar sector, two subtractions are needed
in Eq.~(\eq{DISP}), while only one subtraction is necessary
in the vector sector, because of current conservation.
 
As a consequence of this, our analysis for the scalars does in
principle depend on the gluon and chiral condensates.  The
sensitivity to the value of the gluon condensate turns out to
be very small.  This is not true for the chiral condensate, 
$\langle\psb\psi\rangle$.  We find, in particular, that the
value of the chiral condensate obtained from the vectors, is
too large to satisfy all equations in the scalar sector with
reasonable values for all parameters.  This is consistent with
the fact that this value of the condensate is also on the high side 
in comparison with other estimates. Interestingly, we find that 
the value of the scalar mass is very insensitive to the value 
of the chiral condensate. It depends, however, on the value of $L_8$
(which we used as input). We would
like to emphasize that one should not try to identify this large-$N_c$ scalar
resonance with the broad $\sigma$~\cite{PENN} 
appearing in $\pi-\pi$ scattering, as
the latter is more likely a $\pi-\pi$ bound state~\cite{Oller}, 
whose dynamics is subleading
at large $N_c$, and cannot give rise to a leading large-$N_c$ coupling like
$L_8$. 

To summarize, in all cases we considered, we find a scalar mass between
$\sim 0.7$ and $\sim 1.2$~GeV\cite{comment}, 
where the spread is mostly due to the error in the experimental 
value of $L_8$. 

Finally we use our Eq. (\eq{MS}) in combination with the analysis of
Ref. \cite{Peris-Perrottet-deRafael} to produce a parameter-free determination
of the Gasser--Leutwyler 
$L_i$ couplings in the ${\cal O}(p^4)$ chiral Lagrangian, see 
Fig.~\ref{fig:Lis}. 
The overall agreement for the seven $L_i$ couplings is remarkable.

\bigskip
\noindent {\em Acknowledgements}:

We thank M. Knecht, A. Pich, E. de Rafael, A. Gonzalez-Arroyo and F. 
Yndurain for discussions and comments; and A.A. Andrianov and D. Espriu 
for pointing out a numerical error in a previous version of the manuscript.  

MG thanks the Physics Department
of the University of Washington, where part of this work was carried
out, for hospitality. MG is supported in part by a
Fellowship of the Spanish Government SAB1998-0171 and as a US 
Department of Energy Outstanding Junior Investigator.
SP is supported by the research project CICYT-AEN98-1093 of the Spanish
Government and by TMR, EC-Contract No. ERBFMRX-CT980169 (EURODA$\phi$NE).


\begin{references}
\bibitem{comment} Recently this result has also been obtained independently 
  in Ref. \cite{Andrianov}. Some of the views  of these authors
are however different from ours.
\end{references}
\end{document}